# Generalization of Brillouin theorem for the non-relativistic electronic Schrödinger equation in relation to coupling strength parameter, and its consequences in single determinant basis sets for configuration interactions

Sandor Kristyan

Research Centre for Natural Sciences, Hungarian Academy of Sciences,
Institute of Materials and Environmental Chemistry
Magyar tudósok körútja 2, Budapest H-1117, Hungary, kristyan.sandor@ttk.mta.hu

**Abstract**

The Brillouin theorem has been generalized for the extended non-relativistic electronic Hamiltonian ($H_\nabla + H_{ne} + aH_{ee}$) in relation to coupling strength parameter (a), as well as for the configuration interactions (CI) formalism in this respect. For a computation support, we have made a particular modification of the SCF part in the Gaussian package: essentially a single line was changed in an SCF algorithm, wherein the operator $r_{ij}^{-1}$ was overwritten as $r_{ij}^{-1} \rightarrow ar_{ij}^{-1}$, and "a" was used as input. The case a=0 generates an orto-normalized set of Slater determinants which can be used as a basis set for CI calculations for the interesting physical case a=1, removing the known restriction by Brillouin theorem with this trick. The latter opens a door from the theoretically interesting subject of this work toward practice.

## 1.a. Introduction: The coupling strength parameter

The non-relativistic, spinless, fixed nuclear coordinate electronic Schrödinger equation (SE) for a molecular system containing M atoms and N electrons with nuclear configuration $\{\mathbf{R}_A, Z_A\}_{A=1,...,M}$ in free space is

$$H(a=1)\Psi_k = (H_\nabla + H_{ne} + H_{ee})\Psi_k = E_{electr,k}\Psi_k \qquad (Eq.1)$$

where $\Psi_k$ and $E_{electr,k}$ are the $k^{th}$ excited state (k=0,1,2,...) anti-symmetric wave function (with respect to all spin-orbit electronic coordinates $\mathbf{x}_i \equiv (\mathbf{r}_i, s_i)$) and electronic energy, respectively, as well as the electronic Hamiltonian operator contains the sum of: kinetic energy, nuclear–electron attraction and electron–electron repulsion operators. The most popular calculations [1-2] for deducing important physical properties from Eq.1 as stationary points (geometry minimums and transition states), vibronic frequencies, rotations, van der Waal's interactions, excited states, reaction barriers, reaction heats, etc. are the expensive but (chemical) accurate configuration interactions (CI) method for ground and excited states, and the less accurate but faster and less memory taxing Hartree–Fock self consistent field (HF-SCF) method for a ground state with or without correlation corrections [3-13]. (Of course in practice, the CI and HF-SCF are not the most commonly performed calculations solving the SE, for example, coupled cluster methods (reducing the complexity of CI) and MP2 or DFT calculations (increasing the accuracy of HF-SCF without dramatically increasing the computation time) are much more popular.)

The CI works for any nuclear geometry, while the HF-SCF is only for the vicinity of stationary points where the assumption of the spin-pairing effect is plausible via a single Slater determinant. (Main notations and definitions are summarized at the end to guide reader.)

As indicated in Eq.1, the Hamiltonian can be extended with coupling strength parameter "a" as $H(a)= H_\nabla + H_{ne} + aH_{ee}$ of which only a=1 makes physical sense. The a=0 case mathematically provides a good starting point to solve the very important problem when a=1. In this extended Hamiltonian H(a) the dimensionless coupling strength parameter "a" scales the electron-electron interaction energy, $V_{ee}(a)$, with this simple and precise definition. It has already been shown [14] that, for example, it is capable to correct the HF-SCF energy remarkably well with scaling it a bit below unity. On the other side, the role of coupling strength parameter is also known in the literature which defines the "adiabatic connection (AC) Hamiltonian" [15-22], wherein the Hamiltonian is extended similarly, not purely with operators as above, but in the context of Kohn-Sham (KS) formalism (with the help of one-electron density, etc.). Important to emphasize that, in AC the coupling strength parameter connects the KS system to physical system of interacting electrons (H(a=1)), while in this work it connects an unphysical system (no electron repulsion, H(a=0)) to the system treated at the mean-field Hartree-Fock level and above; (so, the two hierarchies should not be mixed). Below we analyze the behavior of case a=0 with our hierarchy defined here (different than the AC), and its effect on ground and excited states, not only on ground state correlation effects.

The coupling strength parameter [23], we manipulate with, is used in DFT for investigating the "exchange (Fermi) and correlation (Coulomb) hole". In short, these holes are those, which are connected to the error created by the HF-SCF and the DFT based Kohn-Sham (KS) methods [2, 23] try to re-correct during the routine using the one-electron density, or the wave function methods [1] which try to re-correct after the routine using the wave functions. The error stems from the use of one single Slater determinant ($S_0$) to approximate the ground state wave function ($\Psi_0$) when it is effected by the operator $1/r_{ij}$ in the Hamiltonian, and is responsible for the exchange (Fermi hole) error, and correlation (Coulomb hole) error (estimated as $E_{corr} := E_{xc} < 0$) in the calculation of $<\Psi_0|H_{ee}|\Psi_0>$ with approximation $<S_0|H_{ee}|S_0>$. However, in this work we use the coupling strength parameter to investigate the entire term $<\Psi_0|H_{ee}|\Psi_0>$. We note that there is another error stemming from the use of $S_0$ in calculating the kinetic energy, $<S_0|H_\nabla|S_0>$, to approximate $<\Psi_0|H_\nabla|\Psi_0>$, that is about a magnitude less than $E_{xc}$ and has an opposite sign. Furthermore, physicists [2] divide this problem as $E_{corr} := E_x + E_c$, where the $E_x$ accounts for the error stemming from $<S_0|H_\nabla|S_0>$ and $E_c$ from $<S_0|H_{ee}|S_0>$.

**1.b. Changes to the HF-SCF routine: The coupling strength parameter as input**

The standard HF-SCF routine was modified with a few simple program lines, which can be done in any of the existing SCF subroutines: Those lines of the SCF subroutine (particularly in Monstergauss 1981, a very early version of Gaussian package [24]) were modified, which calls for the subroutine to calculate the two and four center integrals (known as K and J integrals [1-2]) for $<S_0|H_{ee}|S_0>$ with particular molecular orbitals (MOs). Simply, the seed term $r_{ij}^{-1}$ was overwritten with $ar_{ij}^{-1}$, and the parameter "a" was programmed as input. Essentially it was a simple modification in one line only as variable:= parameter*variable. The a=1 leaves the operator $H_{ee}$ in full effect in our work in a regular way, while a=0 totally switches it off for our purpose indicated in the title. In this way, this input parameter "a" assumes the role of the coupling strength parameter. (It was just the same as in ref.[14] for a totally different purpose, particularly for correlation calculation, based on a very different theoretical point of view.)

Below we extend the regular notation "HF-SCF/basis" to "HF-SCF/basis/a", wherein the latter indicates the value of the coupling strength parameter "a" used beside the basis set in the standard HF-SCF algorithm, extended with a coupling strength parameter as input described above. The a=1 is the standard HF-SCF/basis for Eq.1 and a=0 is for Eq.2 below, and there can be other values for "a" to manipulate.

With this device, the mode a=0 solved the equation as a counter part of Eq.1:
$$(H_\nabla + H_{ne})Y_k = e_{electr,k} Y_k \qquad (Eq.2)$$
for ground state k=0 (or lowest lying enforced spin multiplicity state) with a Slater determinant for $Y_0$ with HF-SCF/basis/a=0 algorithm. Moreover, even excited states (k>0) can be obtained by HF-SCF/basis/a=0 algorithm for Eq.2, which is definitely not feasible for Eq.1 by only using HF-SCF/basis/a=1 algorithm. The eigenvalue pairs, ($e_{electr,k}$, $Y_k$), - of course - differ from Eq.1, and are notated differently. A very important result we mention, and will be detailed in later work: The $S_0$ obtained by HF-SCF/basis/a=1 performed on Eq.1 is very close to $Y_0$ by HF-SCF/basis/a=0 performed on Eq.2, moreover, calculating $Y_0$ in this way is not restricted to the vicinity of the stationary point. As will be analyzed in section 2, no effect corresponding to correlation effect rises up in Eq.2, the Slater determinant is an adequate form for ground state anti-symmetric $Y_0$, more, for $Y_k$ too, only the problem of not reaching the basis set limit elevates the energy governed by the variation principle. (Recall again that when the single Slater determinant $S_0$ approximates the not single determinant $\Psi_0$, it creates the correlation error beside the basis set error.) Eq.1 has physical sense while Eq.2 does not, but Eq.2 provides a very rich pro-information for Eq.1. Furthermore, the HF-SCF/basis/a=0 calculation for Eq.2 is faster, more stable and less memory taxing in comparison to HF-SCF/basis/a=1 for Eq.1. We describe the mathematical and computing connection between Eq.1 and 2.

**1.c. The aim of this work**

We will demonstrate that, the orthogonal basis set $\{Y_k\}$ from Eq.2 provides simpler Hamiltonian matrix for different level CI calculations in its off-diagonal elements (see Eq.13 later) along with an opportunity to avoid the restriction from Brillouin's theorem, (even the SCF convergence originated from $1/r_{12}$ is eliminated from the algorithm).

**2. The SE (Eq.1, a=1) versus the "totally non-interacting reference system" (TNRS, Eq.2, a=0)**

Similarly to Eq.1, by definition, we ask $Y_k$ for Eq.2 to be anti-symmetric and well behaving (vanishing at infinity and square-integrable), normalized as $<Y_k|Y_k>=1$, and the ground state one-electron density is defined as $\rho_0(\mathbf{r}_1,a=0)= N\int Y_0^* Y_0 ds_1 d\mathbf{x}_2 \ldots d\mathbf{x}_N$, where "TNRS" stands for "totally non-interacting reference system" (name explained below) in analogy to the corresponding anti-symmetric and well behaving $\Psi_k$: $<\Psi_k|\Psi_k>=1$ and $\rho_0(\mathbf{r}_1,a=1)= N\int \Psi_0^* \Psi_0 ds_1 d\mathbf{x}_2 \ldots d\mathbf{x}_N$, respectively. A trivial property is that for N=1, i.e. for H-like atoms (M=1) and molecular frame with one electron (M>1), Eqs.1 and 2 overlap or identical. Furthermore and more importantly, certain theorems for Eq.1 hold for Eq.2 as well, part of it is known in literature, but not summarized as will be below. Most importantly, Eq.2 is a linear partial differential equation, the variation principle holds, and the 1st ("$\rho_0(\mathbf{r}_1,a=0)$ of TNRS defines $Y_0$ and the nuclear frame") and 2nd ("variation principle for $\rho_0(\mathbf{r}_1,a=0)$ of TNRS in the DFT functional stemmed by Eq.2") Hohenberg – Kohn (HK) theorems hold in an analogue sense which will be set out in more quantized form in Eq.9 below.

The density functional for the nuclear–electron term is the simple 100 % accurate - $\Sigma_{A=1,...,M} Z_A \int R_{A1}^{-1} \rho_{trial}(\mathbf{r}_1) d\mathbf{r}_1$, i.e., the same form for both Eqs.1 and 2. Also, the same form for the kinetic term holds for both Eqs.1 and 2 – however, only approximate forms and not exact forms are known as yet. It is also obvious that with the a=0 mode (switching the effect of operator $H_{ee}$ off), Eq.2 can also be treated with the HF-SCF/basis/a algorithm (defined above), and the energetically lowest lying eigenvalue pair ($e_{electr,0}$, $Y_0$) corresponds to ($E_{electr,0}$, $\Psi_0$). Furthermore, $E_{electr,0} \gg e_{electr,0}$ for any molecular system (in stationer or non-stationer geometry), and the large difference mainly stems from the lack of $V_{ee}(a=1)$ when a=0. Moreover, the ground state versus the energetically lowest lying state with an enforced spin multiplicity feature is also the same as in the HF-SCF/basis/a treatment of Eqs.1 and 2 – recall the example of neutral ($1s^2, 2s^2, 2p^2$) carbon atom open shell (triplet, ground state) versus closed shell (lowest lying singlet) states, and so on. However, if spin-spin interaction is not considered via Coulomb repulsion, Hund's rule does not apply for Eq.2 (a=0) itself, for example, but extension and approximation in e.g. Eq.8 below sets it back on the right track, that is, e.g. the triplet has lower energy than the singlet, as in the HF-SCF/basis/a=1 for Eq.1 (p.103 of ref. [1]).

On the other hand, there are major mathematical differences between Eqs.1 and 2 aside from the visible inclusion of or omitted operator $H_{ee}$. For both, operator $H_\nabla$ makes them differential equations, which is generally necessary - philosophically speaking - to describe a physical phenomenon, and for both, $H_{ne}$ defines the nuclear frame, and the molecular system for them. However, operator $H_{ee}$ is very special in Eq.1 in the sense that algebraically it is the "simplest" term, but in contrast, as it has turned out in the history of computational chemistry, it introduces the most difficult effect [25-26] in HF-SCF computation, known as the non-classical Coulomb effect. It is difficult to treat for HF-SCF/basis/a routine via "correlation calculation" or "exchange correlation DFT" devices after or during, respectively [1-2, 23]. $H_{ee}$ operator is responsible for the fact that a single Slater determinant $S_0$ for $\Psi_0$ in Eq.1 is not enough for total accuracy, although in the vicinity of stationary points on the PES, it provides a good approximation, and it can provide many characteristic properties of the ground state eigenvalue. In contrast, a single Slater determinant form is adequate for Eq.2 not only for the ground, but also for excited states, and the HF-SCF/basis/a=0 with basis set limit accurately calculates the eigenvalue pairs ($e_{electr,k}$, $Y_k$) for ground and excited states.

The manipulation with Slater determinants in HF-SCF theory is well established [1], but some textbook properties must be overviewed, since a new aspect is described, that is, we make an allowance for Eq.1 to be replaced by Eq.2. In detail, Eq.2 is $(H_\nabla + H_{ne})Y_k = (-(1/2)\Sigma_{i=1,...,N} \nabla_i^2 - \Sigma_{i=1,...,N}\Sigma_{A=1,...,M} Z_A R_{Ai}^{-1})Y_k = \Sigma_{i=1,...,N} h_i = e_{electr,k} Y_k$, where $h_i \equiv -(1/2)\nabla_i^2 - \Sigma_{A=1,...,M} Z_A R_{Ai}^{-1}$ is the one-electron operator widely used [1] in HF-SCF theory for Eq.1 as well. In the Fock or Kohn Sham equations [1-2, 23] Eq.1 is decomposed to the one-electron equations

$$(h_i + aV_{ee,eff}(\mathbf{r}_i)) \phi_i(\mathbf{r}_i) = \varepsilon_i \phi_i(\mathbf{r}_i) \tag{Eq.3}$$

where $\phi_i(\mathbf{r}_i)$ is the i[th] MO, and technically $\phi_i$ counts the MOs with the idex i, so the notation is reducible from $(h_i, \phi_i(\mathbf{r}_i), \varepsilon_i)$ to $(h_1, \phi_i(\mathbf{r}_1), \varepsilon_i)$ mathematically. $V_{ee,eff}$ is the effective potential from electron-electron repulsion; (other habit [23] is that $H_{ne}$ is shifted algebraically into $V_{ee,eff}$, and called $V_{eff}$, but we do not use that here). Importantly, because $1/r_{ij} \to a/r_{ij}$ change for operator $H_{ee}$ was made in the algorithm, the parameter "a" entered linearly to $V_{ee,eff}$ in Eq.3. $V_{ee,eff}$ in Eq.3 is expressed with the known J and K integrals in HF-SCF theory, or $V_{ee,eff}(\mathbf{r}_i)= \int \rho_0(\mathbf{r}_2,KS)r_{12}^{-1}d\mathbf{r}_2 + V_{xc}(\mathbf{r}_i)$ in Kohn-Sham formalism (the first term is the classical Coulomb term, the second is the non-classical Coulomb term for "exchange-correlation"). $V_{ee,eff}(\mathbf{r}_i)$ is the term where the N equation in Eq.3 is coupled (a=1 or generally a≠0). (The $\rho_0$ depends on $\Psi_0$ which is approximated with a single Slater determinant containing all other $\phi_j$, j=1,...,N and j≠i.) Another property is that MOs are ortho-normal, that is $<\phi_i|\phi_j> = \delta_{ij}$. In Eq.3 the operator seed $1/r_{ij}$ is reduced to the variable, $\mathbf{r}_i$ via performing the integrations, and virtually all equations in Eq.3 depend on one-electron. It is in fact coupled, though virtually not coupled, so the 100% adequate anti-symmetric solution for the equation system in Eq.3 (but not for Eq.1) is a Slater determinant, and this system is known as: "non-interacting reference system" [23], as is well known. System in Eq.3 is commercially programmed [24] by the standard HF-SCF or Kohn – Sham formalism, but if a=0 is set in the input, a special modification in the algorithm for this work, Eq.3 reduces to

$$h_1\phi_i(\mathbf{r}_1) \equiv (-(1/2)\nabla_1^2 - \Sigma_{A=1,...,M}Z_A R_{A1}^{-1})\phi_i(\mathbf{r}_1) = \varepsilon_i \phi_i(\mathbf{r}_1) . \quad (Eq.4)$$

System Eq.4 is the Fock equation system for Eq.2. However, we are at a reduction where a single Slater determinant as anti-symmetric solution is not only 100 % adequate for Eq.4, but also for Eq.2. The reason for this is: that all operators are one-electron operators, the two electron operators with seed elements $1/r_{ij}$ are cancelled by a=0, that is, Eq.4 describes a non-coupled system. For this reason, we call Eqs.2 or 4: TNRS, distinguishing them from the "non-interacting reference system" above. More simply, Eq.4 should not be considered as an equation system containing N equations enumerated by i ($h_i\phi_i= \varepsilon_i\phi_i$), but in fact it is a single eigenvalue equation ($h_1\phi_i= \varepsilon_i\phi_i$). Eigenvalues of Eq.4 are $(\varepsilon_i, \phi_i(\mathbf{r}_1))$ for i=1,2,...∞, the i=1 is the lowest lying state of Eq.4 and it is the lowest lying MO for Eq.2 in its k=0 ground state. The single Slater determinant for Eq.2 is accomplished for N electrons from the eigenvaues of Eq.4, just as in the basic HF-SCF theory. Notice that the HF-SCF/basis/a algorithm (at any "a") is accomplished in such an algebraic way that it keeps the MOs ortho-normal in $s_0(a)$ during the optimizing algorithm, particularly in $S_0=s_0(a=1)$, so for $Y_0=s_0(a=0)$, in great accord that the eigenfunctions of linear Eq.4 are mathematically ortho-normal.

Eq.4 with the value of N and Eq.2 are equivalent, more, it holds for the case too, when mathematically, one needs a symmetric $Y_k$, instead of an anti-symmetric one. The electronic energy of the system in Eq.2 is the sum of energy levels (states of Eq.4 or MOs of Eq.2), generally speaking weighted as populated:

$$e_{electr,k}= \Sigma_{i=1,...} n_i \varepsilon_i , \quad (Eq.5)$$

where $n_i$ is the population of the i[th] energy level: 0, 1 or 2, the lattermost is with opposite spins; if $n_i=0$ for an i-value under HOMO, that $Y_k$ is very likely an excited state, (certain degeneracies can yield ground). Of course, $\Sigma_{i=1,...} n_i = N$ must hold. As it is clear, the excited states of Eq.2 can also be described or accomplished, and a single Slater determinant is a 100 % accurate form for these too, as for ground state. In contrast, a single Slater determinant, e.g. with HOMO $\to$ LUMO excitation for one electron for a low lying excited state is an even worse approximation than the approximation $S_0$ for ground state, both by HF-SCF/basis/a=1 for Eq.1. The analogue of Eq.5 between MO energies ($\varepsilon_i$) and ground state electronic energy ($E_{electr,0}(a\neq0)$) is not held in the context of HF-SCF/basis/a=1 approximation or Kohn-Sham formalism, that is $E_{electr,0}$(HF-SCF or KS/basis/a=1) $\neq \Sigma_{i=1,...} n_i \varepsilon_i$ for deepest possible filling in the single Slater determinant, where $\varepsilon_i$'s are from Eq.3 with a=1 (more generally if a≠0): some cross terms must be subtracted [1]. However, what must be subtracted, that goes to zero if a→0. The definition of restricted (RHF) and unrestricted (UHF) form of Slater determinants also lose their necessity here in Eq.2, while these are a certain handicap in HF-SCF approximation for Eq.1 (to get lower energy from the variation principle allowing more LCAO parameters), recall again that the single determinant is an accurate form of solution for Eq.2, but only an approximate solution for Eq.1. (The $\phi_i$'s are eigenfunctions of Eq.3 if a=0 for Eqs.2 and 4 with k≥0, while they are not when a≠0, particularly when a=1 in Eq.3 for Eq.1 with k=0, only an energy minimization.)

For a value of N and multiplicity 2S+1= 2$\Sigma s_i$+1 in the regular way, the HF-SCF/basis/a=0 algorithm calculates the lowest lying N/2 or (N+1)/2 energy values ($\varepsilon_i$) and MOs ($\phi_i$), the latter with LCAO expansion of the basis set for Eqs.2 or 4. As is known, the regular HF-SCF/basis/a=1 algorithm optimizes an $S_0$ single determinant energetically for Eq.1 keeping MOs of $S_0$ ortho-normal during the optimization. This enforced ortho-normalization is also used in HF-SCF/basis/a=0 mode for Eqs.2 or 4 to obtain $Y_0$, in agreement that eigenvectors of Eq.4 (which are MOs of $Y_0$) is ortho-normal set coming purely from the mathematical nature of Eq.4.

Finally, important for Eqs.2 or 4 is that, 1., the HF-SCF/basis/a=0 (the "a-value modified" commercial HF-SCF/basis/a=1 algorithm) is technically perfectly adequate for solving these equations, and 2., for any molecular geometry on the PES, not only at the vicinity of stationary points (which is a serious restriction of HF-SCF/basis/a≠0 case) the HF-SCF/basis/a=0 gives a mathematically adequate result, stemming from the fact that for a=0, the single determinant is an accurate form of the solution, only basis set error present. The $Y_k$'s of Eq.2 have much better mathematical properties than $S_0$ for $\Psi_0$ of Eq.1, however, $S_0$ has been about ready in practice for a long time, while $Y_k$'s are not, and have to be converted in order to be able to be used for $\Psi_k$'s of Eq.1, this will be outlined below step by step. A Slater determinant is a 100% accurate form of solution ($Y_0 = y_0(a=0)$), that is, there is no Coulomb hole, because there is no electron-electron interaction at all, and no Fermi hole, because the anti-symmetric property is provided 100% by a Slater determinant.

### 3. General functional links between SE (Eq.1, a=1) and the TNRS (Eq.2, a=0) along with coupling strength parameter

An important link between Eq.1 and Eq.2 comes from $<\Psi_0|H_{ee}|Y_0> = <\Psi_0|H - (H_\nabla + H_{ne})|Y_0> = <\Psi_0|H|Y_0> - <\Psi_0|(H_\nabla + H_{ne})|Y_0> = <Y_0|H|\Psi_0> - e_{electr,0}<\Psi_0|Y_0> = E_{electr,0}<Y_0|\Psi_0> - e_{electr,0}<\Psi_0|Y_0>$, where the hermitian property of H (and its three parts) was used. Finally,

$$E_{electr,0} = e_{electr,0} + <\Psi_0|H_{ee}|Y_0>/<\Psi_0|Y_0> \tag{Eq.6}$$

bearing in mind that $\Psi_0$ and $Y_0$ are both anti-symmetric. The analysis and test on 149 (Figure 1) molecular G3 ground state electronic energy [27-28] (with G3 equilibrium geometry, neutral charge ($\Sigma Z_A = N$) molecules and selection $\max(Z_A) \leq 10$) has shown the behavior of $<\Psi_0|H_{ee}|Y_0>/<\Psi_0|Y_0>$ and ratio $(E_{electr,0} - e_{electr,0})/e_{electr,0}$ as a function of molecular frame seeded in operator $H_{ne}$, not reported here, that these two quantities are quasi-constants, which is surprising at first glance, but we call for comparison the virial theorem, namely $(V_{nn} + V_{ne} + V_{ee})/T = -2 = (V_{nn} + v_{ne})/t \equiv (V_{nn} + <Y_0|H_{ne}|Y_0>)/<Y_0|H_\nabla|Y_0>$ holds exactly on atoms, atomic ions and equilibrium/transition state geometry molecules, i.e., the value 2 is invariant on the nuclear frame seeded in $H_{ne}$ and N in molecular systems. (Non-equilibrium molecules obey a slightly more complex virial equation [2], not detailed here.) It is important to emphasize the significant difference between $V_{ee} \equiv <\Psi_0|H_{ee}|\Psi_0> = (N(N-1)/2)<\Psi_0|r_{12}^{-1}|\Psi_0>$, as the electron-electron repulsion energy term in the sum $E_{electr,0} = T + V_{ne} + V_{ee}$, and the corresponding value from Eq.6, $<\Psi_0|H_{ee}|Y_0>/<\Psi_0|Y_0> = (N(N-1)/2)<\Psi_0|r_{12}^{-1}|Y_0>/<\Psi_0|Y_0>$, as the energy increase by electron-electron repulsion between the two Hamiltonians in Eq.1 (electron-electron interaction is on) and Eq.2 (electron-electron interaction is switched off). Because the nuclear-nuclear repulsion energy, $V_{nn}$, is added after the calculation, that is cancelled in the difference in Eq.6, and as a consequence: $E_{total\ electr,0} - e_{total\ electr,0} = E_{electr,0} - e_{electr,0}$. (Notice that the divisor $<\Psi_0|\Psi_0>$ comes up in $V_{ee}$ if it is not normalized to unity, making the algebraic analogy even closer between $V_{ee}$ and $(N(N-1)/2) <\Psi_0|H_{ee}|Y_0>/<\Psi_0|Y_0>$.) A more general expression than Eq.6 is hold between k and k' excited states, coming from the same one-line derivation:

$$E_{electr,k} = e_{electr,k'} + (N(N-1)/2)<\Psi_k|r_{12}^{-1}|Y_{k'}>/<\Psi_k|Y_{k'}>, \qquad (Eq.7)$$

and it forecasts the approximation for ground state (k=k'=0) if one takes the fact that the LCAO coefficients in $S_0$ (for HF-SCF approximate of $\Psi_0$) and $Y_0$ are very close (mentioned above):

$$E_{electr,0} \approx E_{electr,0}(TNRS) \equiv e_{electr,0} + (N(N-1)/2)<Y_0|r_{12}^{-1}|Y_0>. \qquad (Eq.8)$$

On the right hand side of Eq.8 the $(N(N-1)/2)<Y_0|r_{12}^{-1}|Y_0> = 2J - K$, after expanding the determinants, where J and K are the known Coulomb- and exchange integrals, well known in HF-SCF formalism, but here the MOs belong to $Y_0(a=0)$ (and not $S_0(a=1)$). The possible extension of 1st Hohenberg-Kohn (HK) theorem as

$$Y_0(a=0) \Leftrightarrow H_{ne} \Leftrightarrow \Psi_0(a=1) \qquad (Eq.9)$$

also holds, manifesting as Eq.7 with k=k'=0. Further relations are listed in the Appendix 1.

## 4. TNRS ground and excited states $\{Y_k, e_k\}$ from Eq.2 (a=0, k=0,1,2,… in one step using atomic basis set) as an orto-normalized Slater determinant basis set for CI calculations on Eq.1 (a=1, k=0,1,2,…)

The solution for TNRS (Eq.2, a=0) can be technically obtained via the standard HF-SCF algorithm with the device that the coupling strength parameter "a" is programmed as input in a fast and stable way notated as HF-SCF/basis/a=0. Beside the very interesting and important fact that the HF-SCF/basis/a level LCAO parameters for a=0 (TNRS Eq.2) and a=1 (Eq.1) are close to each other (aside from some possible phase factors or degeneracy), the LCAO coefficients of TNRS can be obtained in only one step with HF-SCF/basis/a=0 algorithm for Eq.2, irrespective of system size (i.e. no steps to convergence is necessary when a=0, only a one step eigensolving). In contrast, the HF-SCF/basis/a=1 LCAO coefficients of a real system (or the mathematical a≠0 cases) can only be obtained through many steps; operator $H_{ee}$ is responsible for this, and the number of steps dramatically increases with the number of electrons (N), more precisely, with system size. A finer detail is that if a convergence problem arises from system size when a=1 and large N, the breakdown never happens in the first step, but usually much later, at least this is our experience in practice. We mention again that for Eq.2 (a=0) the solutions $Y_k$ (k=0,1,2,...) have an exact Slater determinant form, while for other coupling strength parameter values (a≠0) the form of solutions, $y_0(a)$, are not single determinant forms creating the correlation effect if single determinant ($s_0(a)$) is used to approximate $y_0(a)$. It is also known that for real (a=1) systems the HF-SCF/basis/a=1 approximation with basis set and correlation error, yields physically trustable MOs up to LUMO, though higher MOs must be considered with caution. In more detail, estimation HF-SCF/basis/a≠0 for ground state provides the virtual orbital LUMO (≡LUMO+1) as a side product, which can be used as a weak estimation of the first excited state, but LUMO+2, LUMO+3, etc. cannot be used for this purpose, alone they would be mathematically very weak to estimate the eigenvalues of operator $H_\nabla+H_{ne}+aH_{ee}$, and only good for constructing orthonormal basis sets for CI calculations in a next step (the manipulation $s_0(a) \to s_{0,p}^q(a)$, etc.). However, the HF-SCF/basis/a=0 MOs are mathematically correct up to N/2 or (N+1)/2 and over, suffering from basis set error only. (These LUMO+1,2,... orbitals are created as a side product over the lowest lying N/2 or (N+1)/2 MOs, as well as require high enough AOs to be used in the basis set, and more importantly, they can have physical meaning after a further plausible process which can transfer them from Eq.2 to Eq.1.) In ensemble DFT for excited states using minimal basis can be found in ref. [29].

The manipulation with the Slater determinants in CI theory is well established [1], but some textbook properties must be overviewed. To obtain ground ($Y_0$) and excited ($Y_k$) states of Eq.2 via the HF-SCF/basis/a=0 algorithm is as follows: The ground state $Y_0$ can be calculated as described above; it solves Eq.4 first for some $\phi_i$ states, and sets up the ground state ($Y_0$) for Eq.2. With e.g. side product LUMO which satisfies Eq.4 as the other MOs, the excited states ($Y_k$) can also be set up. For excited states one has to provide basis set adequate for higher $\phi_i$ (i > N/2 or (N+1)/2) states: for example, for the close to neutral (charge -1, 0 or 1) small hydrogen-fluorid or similar kinds of molecule, e.g. the STO-3G or 6-31G*, etc. basis sets include the 1s, 2sp and in the latter the 3spd AOs [24] for approximating $\Psi_0$ in Eq.1 using HF-SCF/basis/a=1 algorithm (suffering with basis set and correlation error), these basis sets were worked out for these kinds of systems. However, these basis sets can also be used for HF-SCF/basis/a=0 algorithm (causing basis set error only) for approximating $Y_0$ (Eq.2) or $\phi_i$ (Eq.4) being the 2p the highest occupied AO in fluoride participating in the hydrogen-fluoride molecule. Calculating higher TNRS excited states for hydrogen-fluoride via Eqs.2 or 4, the 4spdf, 5spdfg, etc. AOs are also necessary, depending how high the values k or i are targeted in $Y_k$ (or $\phi_i$). (This must also be provided if (a different kind of) CI algorithm is performed on Eq.1.) The HF-SCF/basis/a generated {$Y_k(a=0)$} determinant basis set can be used for CI calculations, as the {$S_k(a=1)$} in practice, the linear algebra is exactly the same, but the algebraic forms do differ a slightly, of course, and this will be elaborated upon next.

In calculating $Y_k$ of Eq.2, one has to apply the standard way of linear algebra for energy Hamiltonian [1] which requires to compute the matrix elements $<b_i|h_1|b_j>$ for i,j=1,2,…,K, where {$b_1(\mathbf{r}_1)$, $b_2(\mathbf{r}_1)$,…$b_K(\mathbf{r}_1)$} is an adequate, atom-centered AO basis set [24]. The K eigenvalues (MO energies) and eigenvectors (wave functions) of this KxK Hamiltonian matrix approximates the lowest lying K eigenstates: the orbital energy values, {$\varepsilon_i$}$_{i=1..K}$, and ortho-normal wave functions, {$\phi_i(\mathbf{r}_1)$}$_{i=1..K}$, of Eq.4. This is what we call a one step algorithm, because the eigensolver is used only once. Now a=0, but recall that in HF-SCF/basis/a≠0 algorithm every step after the initial estimation needs eigensolver, what HF-SCF or KS do during a typical (a=1) SCF device [1-2]. As a result, the $\phi_i(\mathbf{r}_1)$ wave functions (of Eq.4 and the so-called MOs of Eq.2) are expressed in LCAO in the basis set chosen, and are ortho-normal required by Eqs.2 and 4. The eigenstate energy values {$e_{elekt,k}$}$_{k=0,1,..L-1}$ along with the set of single Slater determinant wave functions {$Y_k$}$_{k=0,1,..L-1}$ can be accomplished systematically by mixing $\phi_i$ (i= 1.2,…,N/2 or (N+1)/2, …K) as in the standard algebra with a Slater determinant for case a=1, i.e. for {$S_k$} [1], but there the corresponding equation is not as simple. The L=(2K)!/(N!(2K-N)!), since all $\phi_i$ can be non-, singly- or doubly (and oppositely) occupied by α or β spins.

What is important is the orthogonal property

$$<\phi_i(\mathbf{r}_1)|\phi_j(\mathbf{r}_1)> = <Y_i(\mathbf{x}_1,…,\mathbf{x}_N)|Y_j(\mathbf{x}_1,…,\mathbf{x}_N)> = \delta_{ij} \quad (Eq.10)$$

where obviously, the bra-ket (<|>) integration means 3 and 4N dimensions, respectively. Normalization $N<Y_i|Y_i> = \int\rho_i(\mathbf{r}_1,a=0)d\mathbf{r}_1 = N$ also holds, as a conventional definition for i$^{th}$ excited state. Property in Eq.10 for orbital set {$\phi_i(\mathbf{r}_1)$} and determinant set {$Y_k$} comes from the hermitian and linear nature of the operators in Eq.4 and Eq.2, respectively, that is,

$$\varepsilon_j\delta_{ij} = <\phi_i|h_1|\phi_j> = <\phi_j|h_1|\phi_i> = \varepsilon_i\delta_{ji} \quad (Eq.11)$$

and,

$$e_{electr,j}\delta_{ij} = <Y_i|H_\nabla + H_{ne}|Y_j> = <Y_j|H_\nabla + H_{ne}|Y_i> = e_{electr,i}\delta_{ji} \quad (Eq.12)$$

as well as for basis set elements $<b_i|h_1|b_j> = <b_j|h_1|b_i>$. Now, the subject in section 2 is generalized from ground state to excited states determinants. The normalization in Eq.10 is just a matter of using a proper constant multiplier with $\phi_i$ or $Y_k$. The anti-symmetric orto-normalized Slater determinant basis set {$Y_k$} from Eq.2 (a=0) via HF-SCF/basis/a=0 algorithm which is also complete i.e., any anti-symmetric function for interchanging any ($\mathbf{x}_i$, $\mathbf{x}_j$) pair in 4N dimensional ($\mathbf{x}_1$,…, $\mathbf{x}_N$) space that can be expanded with them, can be used for solving Eq.1 for ground and excited states with determinant expansion of $\Psi_k$, like the ortho-normal {$S_k$} set from HF-SCF/basis/a=1 in practice.

For an LCAO estimation by HF-SCF/basis/a=0, the main step is the above diagonalisation of matrix $<b_i|h_1|b_j>$. Of course, the computation time increases for this one main step with the size of Hamiltonian matrix, that is, with system size. The program to solve the eigenvalue problem $<b_i|h_1|b_j>$ is straightforward, but an HF-SCF/basis/a=0 algorithm (from commercial programs modified with the coupling strength parameter "a" as input) can conveniently be used.

## 5. Tricking the HF-SCF/basis/a=0 algorithm to obtain excited states, $\{Y_k\}$

If higher states than $Y_0$ are required, one can still use the existing HF-SCF/basis/a=0 codes, but one has to do a trick with changing the charge of system because only LUMO+1 or LUMO+2 come out purely as a side product, and no higher excited states. That is, simply increasing N on the same nuclear frame $\{R_A,Z_A\}_{A=1,...,M}$ encapsulated in $H_{ne}$ by using the correct multiplicity, although the latter allows greater freedom than in the case a=1; a simple and typical demonstration of this follows with hydrogen-fluorid molecule (MP2(full)/6-31G* geometry, $E_{total\ electr,0}$(MP2 level)= -100.1841 Hartree [24]), the HF-SCF/STO-3G/a=1 for Eq.1 yields:

```
1CLOSED SHELL SCF, NUCLEAR REPULSION ENERGY IS 5.099731703 HARTREES
0CONVERGENCE ON DENSITY MATRIX REQUIRED TO EXIT IS  1.0000D-05
0 CYCLE    ELECTRONIC ENERGY      TOTAL ENERGY    CONVERGENCE   EXTRAPOLATION
    1       -103.453458282        -98.353726579
    2       -103.658442376        -98.558710673    4.81239D-02
    3       -103.671344215        -98.571612512    1.00099D-02
    4       -103.671920720        -98.572189017    2.13556D-03    4-POINT
    5       -103.671934950        -98.572203247
    6       -103.671950402        -98.572218699    5.80744D-06
0AT TERMINATION TOTAL ENERGY IS      -98.572219   HARTREES
1MOLECULAR ORBITALS                      5 OCCUPIED MO
                          1         2         3         4         5         6
     EIGENVALUES---   -25.90153  -1.46601  -0.58015  -0.46365  -0.46365   0.61156

  1  1   F   1S       0.99472  -0.24986   0.08063   0.00000   0.00000   0.08298
  2  1   F   2S       0.02247   0.94095  -0.42420   0.00000   0.00000  -0.53979
  3  1   F   2PX      0.00000   0.00000   0.00000   0.28444  -0.95869   0.00000
  4  1   F   2PY      0.00000   0.00000   0.00000   0.95869   0.28444   0.00000
  5  1   F   2PZ     -0.00283  -0.08462  -0.70026   0.00000   0.00000   0.82101
  6  2   H   1S      -0.00558   0.15494   0.52694   0.00000   0.00000   1.07402
```

while the HF-SCF/STO-3G/a=0 for Eq.2 yields:

```
1CLOSED SHELL SCF, NUCLEAR REPULSION ENERGY IS 5.099731703 HARTREES
0CONVERGENCE ON DENSITY MATRIX REQUIRED TO EXIT IS  1.0000D-05
0 CYCLE    ELECTRONIC ENERGY      TOTAL ENERGY    CONVERGENCE   EXTRAPOLATION
    1       -151.075395174       -145.975663471
    2       -152.831334744       -147.731603041   0.00000D+00
0AT TERMINATION TOTAL ENERGY IS     -147.731603   HARTREES
1MOLECULAR ORBITALS                      5 OCCUPIED MO
                          1         2         3         4         5         6
     EIGENVALUES---   -40.59236  -9.55517  -8.81672  -8.72571  -8.72571  -4.49671

  1  1   F   1S       1.00121   0.23152   0.08800   0.00000   0.00000   0.03901
  2  1   F   2S      -0.00549  -1.03159  -0.35933   0.00000   0.00000  -0.40485
  3  1   F   2PX      0.00000   0.00000   0.00000  -0.03804  -0.99928   0.00000
  4  1   F   2PY      0.00000   0.00000   0.00000  -0.99928   0.03804   0.00000
  5  1   F   2PZ      0.00024   0.44410  -0.94971   0.00000   0.00000   0.26910
  6  2   H   1S       0.00188   0.20530  -0.09439   0.00000   0.00000   1.18497
```

MOs 1-5 are occupied pair-wised with opposite spins by the N=10 electrons in a ground state, the 5$^{th}$ is the highest occupied MO (HOMO) and the 6$^{th}$ MO is the virtual lowest unoccupied MO (LUMO) in both lists. The LUMO in HF-SCF approximation can be handled relatively easily for qualitative discussions, but one must be careful in a quantitative argument. The 6th, unoccupied MO (LUMO) also has a similar LCAO coefficient in the case of a=1 and a=0, just as the other 1-5th MOs aside from phase factors. The rod shaped hydrogen-fluorid defines direction, it was positioned along the z axis, and as a consequence it is reflected in the approximate atomic 2p like MOs (MO 3, 4 and 5 in both cases a=1 and 0), and in relation to our paper, now the order number of the MOs are the same in comparison to a=1 and a=0, but sign change happened, e.g. in the 2nd MO: sgn(0.94095) vs. sgn(-1.03159). An important feature is, that the TNRS (HF-SCF/basis/a=0) already indicates the bond (now by shifting the LCAO from -1 value) in the same way as the regular HF-SCF/basis/a=1 does, the latter is well known. Recall that, both (a=0 and 1) yield 2 equivalent $p_x$ and $p_y$ as well as a different $p_z$ along the bond.

In this example, the HF-SCF/STO-3G/a=0 calculation for Eq.2 or 4 was processed in neutral ($\Sigma Z_A$=N=10) and singlet (1+2$\Sigma s_i$=1) mode, as usual. It calculates the lowest lying N/2= 5 doubly occupied molecule orbitals (MOs). That is, this routine calculates $Y_0$ for Eq.2 as output by listing the five MOs and their energy eigenvalues belonging to $Y_0$, which are, at the same time, the wave functions $\{\phi_1,..,\phi_5\}$ and eigenenergies $\{\varepsilon_1,..,\varepsilon_5\}$ of Eq.4. Calculating higher MO or $\phi_i$, one simply has to increase the number of electrons, e.g. adding -1 charge to the molecule (N=11), and using correct multiplicity (here 1+2$\Sigma s_i$= 2). This HF-SCF/STO-3G/a=0 calculation for Eq.2 or 4 yields exactly the same LCAO coefficients and energy eigenvalues as for the neutral (N=10) hydrogen-fluoride, because it is the TNRS (a=0) calculation; only instead of 5 doubly occupied MO, there are 5 doubly occupied MOs plus 1 singly occupied 6th MO, which is the $\phi_6$ of Eq.4. As expected, the LUMO virtual orbital (N=10, 6th MO) and the HOMO (N=11, 6th) uppermost occupied MOs are exactly the same, which is definitely not true if it is not a TNRS i.e., a≠0 wherein all $\{\varepsilon_i,\phi_i\}$ changes if N changes. (The $E_{total\ electr,0}$ is different, of course, -147.731603 hartree (N=10) and -152.228311 hartree (N=11) now, because Eq.5 has one more term.) To calculate the first $\{\phi_1,..,\phi_8\}$ states with orbital energies $\{\varepsilon_1,..,\varepsilon_8\}$ of Eq.4, one has to set up N=15 (charge -5) and multiplicity 2 or N=16 (charge -6) and multiplicity 1 in the HF-SCF/basis/a=0 routine for Eq.4, both yield the same results, only the occupation of HOMO $\phi_8$ is different: $\phi_8^1$ in the former and $\phi_8^2$ in the latter. And of course, the first 5 states are the same again as for N=10 and 11 above, however, one must be aware of the basis set chosen at this point, see next paragraph. As the HF-SCF/basis/a=1 names the LUMO+1,2,… as "virtual" MOs, this tricking above with N could have the name "virtual" N. The fact that a [HF]$^{-6}$ for the calculation of TNRS states by HF-SCF/basis/a=0 occupied up to i=8 is not a stable molecule in nature in any geometry is irrelevant now, because this is a trick only to obtain higher TNRS states of Eq.2. If the first two virtual MOs (LUMO+1,+2) are enough, the original N does not have to be increased. More, if LUMO+1 is enough to stop to create $\{Y_k\}$ truncated basis set, since LUMO+1 is always on printout list, N does not even have to be increased at all (to N+1 or N+2, see list above); for even N it allows to generate 1+N/2 singlet only-doubly-MO-occupied $Y_k$, and even more (branches including "1+N/2 over 2") singlet (not-only-doubly-MO-occupied) and triplet $Y_k$. It is like the standard single determinant basis set generating combinatorial algorithm [1] with the known HF-SCF/basis/a=1, but now with TNRS (HF-SCF/basis/a=0). Notice also that, as shown in the list above for hydrogen-fluorid, the case a=1 needs many steps in HF-SCF to converge, but case a=0 needs in fact only one. Recall also that by Brillouin theorem, the CI basis set generation by HF-SCF/basis/a=1 for Eq.1 restricts that one must not stop at single excitation, double excitation is necessary, while CI basis set generation by HF-SCF/basis/a=0 for Eq.1 removes this restriction, detailed below, an important practical advantage beside the theoretically interesting generalization of Brillouin theorem along coupling strength parameter (a), the subject of this work.

The neutral (N=10) HF-SCF/basis/a=0 ground state hydrogen-fluoride (wherein the $\{\phi_1,..,\phi_5\}$ are doubly occupied and all levels above are empty) the $\{\phi_6, \phi_7, \phi_8\}$, i.e., the LUMO+1,+2 and +3 can be used e.g. for single-double-excitation-CI, wherein the electron-electron interaction is taken care of by the eigenvalues adequately large CI matrix to estimate the desirable and the physically plausible lowest lying solutions of Eq.1 better than Eq.8, see below for more details. However, attention has to be paid to the basis set. In this example for $\phi_7$ and higher states, an STO-3G basis set (which is defined originally for commercial Gaussian package [24] for close to neutral molecules and is widely used in practice), containing 6 elements (see the 6 rows of LCAO coefficients for hydrogen-fluoride) is not adequate for calculating a progressively negatively charged molecule, because only six independent vectors of LCAO coefficients exist in the space spanned by the STO-3G basis. Even the quality of the first virtual state wave function $\phi_6$ with orbital energy $\varepsilon_6$ is questionable with STO-3G basis, because altogether, the AO of $1s^1$ of H atom and $1s^2 2s^2 p^5$ of F atom form five saturated (doubly occupied) MOs, that is, N=10 electrons on $\{\phi_1,..,\phi_5\}$ states. It means that a minimal basis set must contain higher AOs, i.e., basis functions 2sp on H and 3spd on F for an adequate description of $\phi_6$ and linear independency for $\phi_7$ in this calculation - the only known criteria for this manner of choosing a basis set. Recall the old textbook example of $H_2O$ which is mistakenly predicted to be linear, if only the 1s and 2s AOs are in a basis set lacking the 2p orbitals. Furthermore, large and high quality basis sets are available [24]; the simplest STO-3G has been chosen here for easy discussion. Larger size, higher quality bases sets yield more accurate orbital wave functions and energies, of course, but at the cost of longer computation.

We emphasize that generating the set $\{Y_k\}$ with a HF-SCF/basis/a=0 algorithm is simpler, faster (one step), more effective (larger k) and more convenient than generating set $\{S_k\}$ with an HF-SCF/basis/a=1, although, the author knows perfectly well that the latter is effectively used and widely tested in practice. Additionally, as outlined above, in the case of a=1 the LUMO and up bear the properties of $S_0$, passing it to the ortho-normal basis set for the CI calculation it generates, but only $S_0$ has a really close relationship to Eq.1, while on the other hand, in the case of a=0 all $Y_k$ are the solution of Eq.2, which is is a certain pro-equation of Eq.1.

**6.a. The ground and excited states of SE (Eq.1, a=1) using the nuclear frame generated orto-normalized Slater determinant basis set {$Y_k$} from TNRS (Eq.2, a=0) for different levels of CI calculation**

Above, we have demonstrated that the LCAO coefficients of TNRS obtained by HF-SCF/basis/a=0 algorithm for Eq.2, suffering only from basis set error, but the correct form of wave function, the single determinant is used, and of an HF-SCF/basis/a=1 approximation for Eq.1 suffering not only from basis set error but the lack of correlation estimation too, stemming from the use of the inadequate single determinant form are close to each other - on the same basis level of course. To obtain more accurate ground and lowest lying excited states, or to simply obtain the ground state more accurately than the HF-SCF/basis/a=1 algorithm can provide, the different levels of the CI methods can be used, using excited Slater determinant N-electron basis functions beside the one in the ground state in the many determinant expansion for $\Psi_k$ of Eq.1. In the literature the basis set {$S_k$} is obtained from $S_0$ and LUMO+1, 2 of HF-SCF/basis/a=1 algorithm [1]. Here we will not go into the extensive literature of correlation calculations on the HF-SCF method (post HF-SCF methods) [1] and during HF-SCF method (KS formalism [2]) or general DFT methods [2] nor the different versions of CI methods [1]. We shall only mention that generally, CI calculations are the most accurate and most time consuming methods, as well as it is *ab initio*, i.e., they do not use empirical parameters, only physical constants like Planck's constant (h), etc.. Its error can come only from the lower quality of a generally AO basis set {$b_k(\mathbf{r}_1)$} for MOs to expand with LCAO coefficients, and the truncation of determinant basis set (in the range of simpler to more complicated, i.e., full CI). In practice {$S_k$} is used, but the soon to be introduced {$Y_k$} can also be used, and in principle, it is also a mathematically correct way to obtain wave functions $\Psi_{k'}$ via expansion with them. (We mention the plane wave expansion (PWE) vs. the LCAO expansion, useful for describing crystals vs. molecules, though not detailed here.)

Using Eqs.10-12, the standard way of expanding [1] anti-symmetric wave functions $\Psi_{k''}$ of Eq.1 using the basis set {$Y_k$} from Eq.2 is the linear combination $\Psi_{k''}= \Sigma_k c_k(k'')Y_k$, where beside the TNRS ground state $Y_0$, the TNRS $Y_k$ with k> 0 are the single, double, triple, etc. excited N-electron Slater determinants. This determinant expansion and treatment of the called multi-determinant representation of the exact k''$^{th}$ excited state wave function $\Psi_{k''}$ (k''=0,1,…) of Eq.1 is a very well known procedure, described in ref.[1], using the {$S_k$} determinant basis from an HF-SCF/baisis/a=1 algorithm. However, we now introduce the use of the TNRS basis set {$Y_k$} from Eq.2 via HF-SCF/basis/a=0 algorithm and detail its calculus here. We draw the attention to the fact that the {$Y_k$} from solving Eq.2 is a nuclear frame ($H_{ne}$) generated basis set, containing strong and pure information about the nuclear frame via Eq.2. The determinant set, {$S_k$}, is generated from $S_0$ by HF-SCF/basis/a=1 which approximates (energy minimization) the ground state $\Psi_0$ of Eq.1. $S_k$'s are worse approximations for excited states $\Psi_k$ than the $S_0$ for $\Psi_0$, but the HF-SCF/basis/a=0 generated Slater determinants are the correct ground and excited state solutions of Eq.2 as an alternative. $S_0$ suffers from basis set error and correlation effect, while $Y_k$ suffers only from basis set error, but $S_0$ is just about physically plausible, while $Y_k$ must be converted to have any physical meaning, – notice that $Y_k$ invokes ground and excited states in contrast to the mere ground state $S_0$. By the principles of linear algebra, changing basis set should not be a problem, mainly from the point of view discussed above, inasmuch as LCAO coefficients do not vary greatly with the value of the coupling strength parameter "a" namely, between a= 0 and 1. What is important is that both, $\{S_k\}$ and $\{Y_k\}$, are ortho-normal N-electron determinant basis set for CI, and both are adequate to expand $\Psi_{k''}$ wave functions with them, they behave as basis set change in relation to each other.

The particular generation of basis sets is as follows. Both, $Y_0$ and $S_0$, have been demonstrated for the hydrogen-fluoride molecule above, the 5 occupied and 1 unoccupied (LUMO, the 6$^{th}$) are listed. The 6$^{th}$ MO in these examples is ready to be used to generate single excited HF-SCF/STO-3G Slater determinants, $\{S_k\}$ as $S_0 \rightarrow S_k \equiv S_{0,i}^6$, etc. with i=1,2,…5 in CI calculations for Eq.1 (a=1), but the $\{Y_k\}$ as $Y_0 \rightarrow Y_k \equiv Y_{0,i}^6$, etc. (i.e. of Eq.2, a=0) determinant basis set is also ready to be used in CI calculations for Eq.1 (a=1), too. The ground-, single, doubly, etc. excited states are textbook routines [1] for generating excited elements in the determinant basis set $\{Y_k\}$ obtained from HF-SCF/basis/a=0 or $\{S_k\}$ from $S_0$ from the HF-SCF/basis/a=1 algorithm. We discuss here $\{S_k\}$ and $\{Y_k\}$ parallel to show the position of $\{Y_k\}$ in the business.

Standard linear algebra provides the set of eigenstates $\{\Psi_{k''}, E_{electr,k''}\}$ of Eq.1 by expanding $\Psi_{k''}$ in basis set $\{Y_k\}$: one must diagonalize the Hamiltonian matrix $<Y_{k'}|H_\nabla+ H_{ne}+ H_{ee}|Y_k>$ as the second main step, just like the first main step, the diagonalization of $<b_i|h_1|b_j>$ for the set of eigenstates $\{Y_k, e_{electr,k}\}$ of Eq.2 (HF-SCF/basis/a=0) with tricking (virtual) N as described above if necessary. Using the properties of Eq.2 as

$$<Y_{k'}|H_\nabla+H_{ne}+aH_{ee}|Y_k> = e_{electr,k}<Y_{k'}|Y_k> + a<Y_{k'}|H_{ee}|Y_k> , \quad (Eq.13)$$

where we have extended it with the coupling strength parameter "a" to be more general. With Eq.10, the diagonal elements (k'=k) reduce to

$$<Y_k|H_\nabla+H_{ne}+aH_{ee}|Y_k> = e_{electr,k} + a(N(N-1)/2)<Y_k|r_{12}^{-1}|Y_k> . \quad (Eq.14)$$

Importantly, the generalization of Eq.8 has been obtained (using Eq.7) making the link between case a=0 and 1 for ground (k=0) and excited (k>0) states as:

$$E_{electr,k} \approx E_{electr,k}(TNRS) \equiv e_{electr,k} + (N(N-1)/2)<Y_k|r_{12}^{-1}|Y_k> . \quad (Eq.15)$$

Eq.15 is based again on the knowledge that LCAO parameters do not vary greatly between $\Psi_k$ and $Y_k$, although it was only exhibited above for $\Psi_0$ and $Y_0$. However, Eq.15 gives exact diagonal elements, as well as again being the first approximation from the Rayleigh-Schrödinger perturbation theory as was used above for k=0. We emphasize that approximation Eq.15 has to be regarded with caution until it is fully tested, its k=0 bottom case in Eq.8 is at least tested here, displayed on Figure 2, see caption. HF-SCF/basis/a=1 mode provides ground state $S_0$ and $E_{electr,0}$(HF-SCF), like the weaker HF-SCF/basis/a=0 extended with Eq.8, but HF-SCF/basis/a=0 mode can provide more, the excited states in Eq.15 as a simple estimation – of course all these can be continued with CI calculations or DFT corrections for higher accuracy. Furthermore, if one orders $E_{electr,k}$ as usual as $E_{electr,k} \leq E_{electr,k+1}$ for k=0,1,2,…, it must be proved that $Y_k$ belongs to $\Psi_k$, i.e., that no energy value switch or cross correspondence is in a (k,k+1) pair, for example, which is a plausible hypothesis, ($S_k$ is problematic in this respect, because one has to stop around LUMO). If "a" differ from zero by small (infinitesimal) value $\delta a$, the $Y_k$ and $y_k(a=0+ \delta a)$ obviously belong to each other in relation to a running k. (The "$\leq$" necessary in relation to energy, the "<" is not enough, because TNRS can remove degeneracy gaps, manifesting in Hund's rule extended with "a", moreover, $\Psi_k$ characteristically can have degeneracy.) Recall that in practice, based on the HF-SCF/basis/a=1 algorithm, i.e., the standard HF-SCF/basis method, calculating an excited state is more problematic than the ground state, that is, the LUMO+1 and up has to be regarded with caution. In contrast, the simple right hand side of Eq.15 can be calculated with the HF-SCF/basis/a=0 and the trick with the molecular charge as was discussed above is plausible for any k as a first approximation, but needs refinement.

The quality of estimation in Eq.15 will be tested in a later work. However, we mention without detailing that the Moller-Plesset (MP) first correction for state k of Eq.2 to Eq.1 is just Eq.15. In this way, as k=0 is corrected with states k=1, 2, etc., similarly, the k>0 can be corrected with 0,1,2,…,k+1, k+2, etc. states, that is, with the MP2 analogue $|<Y_k|r_{12}^{-1}|Y_{k,pq}^{rs}>|^2/(\varepsilon_p+ \varepsilon_q- \varepsilon_r- \varepsilon_s)$ terms, wherein r and s can mean de-excitation too. The latter is another justification for Eq.15 beside the idea of the Hamiltonian matrix, so Eq.15 is a plausible approximation as a first guess. Recalling that for a symmetric matrix the sum of its eigenvalues is equal to its trace, i.e., the sum of the diagonal elements, the diagonal elements in Eq.14 can be summed up for k yielding another relationship. (Notice that when e.g., MP2 terms correct the energy, the factors $(\varepsilon_p+ \varepsilon_q- \varepsilon_r- \varepsilon_s)^{-1}$ have positive signs in de-excitation and negative in excitation determinants for k>0, while it is always negative when k=0.)

Applying Eq.10 again, the off-diagonal elements (k'≠k) reduce to

$$<Y_{k'}|H_\nabla+H_{ne}+aH_{ee}|Y_k> = e_{electr, k \text{ or } k'}<Y_{k'}|Y_k> + a<Y_{k'}|H_{ee}|Y_k> =$$
$$= a(N(N-1)/2)<Y_{k'}|r_{12}^{-1}|Y_k> \quad \text{if } k' \neq k, \quad \text{(Eq.16)}$$

that is, the off diagonal elements contain the electron-electron interaction only, which means qualitatively, that the purely Coulomb operator off-diagonal elements correct the deviations via diagonalization that the diagonal elements (Eq.14 or Eqs.8 and 15) make as error to finally obtain the desired $E_{electr,k}$. Notice again that the ($Y_k$, $e_{electr,k}$) pairs (k=0,1,2,…) in Eq.15 can be obtained by one main step (eigensolving $<b_i|h_1|b_j>$ by a HF-SCF/basis/a=0 algorithm) after calculating the integrals in Hamiltonian matrix elements and following with the diagonal coulomb integrals $<Y_k|r_{12}^{-1}|Y_k>$. The latter comes into being after mixing up $Y_k$, irrespectively of the size of the molecular system. (Irrespectively means that this is one main step in the algorithm, but of course, the size of Hamiltonian matrix increases by minimal or higher quality basis sets reflecting the molecular frame and N.) Anyway, Eq.15 with e.g., an effective and quick DFT correction would greatly cancel out the robust CI method for excited states.

In practice, where the set of Slater determinants in use is not the $\{Y_k\}$ by HF-SCF/basis/a=0, but $\{S_k\}$ by HF-SCF/basis/a=1, the off-diagonal elements corresponding to Eq.16 contain orbital energies of MOs too (see Table 4.1 on p.236 of ref.[1]), while the corresponding orbital energies (the $\varepsilon_k$'s in Eq.4) are missing in Eq.16. In this mathematical formalism, those $e_{electr,k}$ are enough to be included in Eq.14 only. The latter may indicate that there is strong correspondence and order between indexing in $\{Y_k\}$ and $\{\Psi_k\}$ mentioned above. It is obvious between $Y_0$ and $\Psi_0$, but true for $Y_k$ and $\Psi_k$ for any k in which the degeneracy can slightly modify.

Not surprisingly, the matrix in Eq.13 is diagonal for a=0, because the set of wave functions $\{Y_k\}$ is expressed trivially with itself, and anyway, in this case Eq.1 reduces to Eq.2. Analogously, Eq.13 is diagonal for a=1 (more generally for any "a") if the orto-normalized eigenfunction set $\{\Psi_k\}$ from Eq.1 (more generally $\{y_k(a)\}$) is used as basis set, however this basis set is the one we are looking for (now for a≠0 and in practice particularly for a=1), the most important set in quantum mechanics when a=1, known analytically for only one-electron atoms, M=N=1, – for which, actually, the value of "a" is irrelevant and Eqs.1 and 2 overlap via N=1. If the off-diagonal elements (Eq.16) are neglected, the matrix in Eq.13 diagonalizes to Eqs.14 or 15, so approximations Eqs.8 and 15 are theoretically supported now via linear algebra, the word "approximation" cannot be over emphasized, see $E_{total\ electr,0}(G3)-E_{total\ electr,0}(TNRS)$ plotted with a solid line in Figure 2, which is remarkable but, far beyond chemical accuracy. If Eq.16 is set to be zero on the right, the coupling strength parameter can be used to empirically re-correct this simplification via an empirical value (shifting "a" back from a=1), that is, with slightly less than unity, see e.g. ref.[14], yet more sophisticated corrections (correlation) are necessary (MP theory was mentioned as a possibility above, for example,), if one wants to avoid the eigensolving (CI method) Eq.13.

The Coulombic terms $<Y_{k'}|r_{12}^{-1}|Y_k>$ in Eqs.14 and 16 generate many products, although the orthogonality of MO set $\{\phi_i\}$ from Eq.4 generating $\{Y_k\}$ makes many cancellation as it is well known (see again Table 2.4 on p.70 or Table 4.1 on p.236 of ref.[1]) so, calculating these matrix elements is in fact not difficult though, time consuming. One important thing must be mentioned. The spin related properties and manipulations are exactly the same in the case of $\{Y_k\}$ as in the case of $\{S_k\}$, since only the LCAO coefficients differ somehow, but the size of the determinants and the topology (energy ladder) is the same, and must be taken into account in the same way. Eq.15 is a lucky form, because only one single determinant is involved, so its spin state is obvious. The only problem is that Eq.15 is not accurate enough, so a correction from DFT for excited states would be useful using e.g., $\rho_k(\mathbf{r}_1)$ from $Y_k$. However, if one wants to use a kind of CI correction, i.e., use off diagonal elements in Eq.16 for correction, then the spin situation must be taken into account. The spin states for Eqs.2 and 4 is as follows: The Hamiltonian does not contain any spin coordinates in Eqs.1-2 and hence both, $S_{op}^2$ and $S_{op,z}$ total spin operators commute with it:

$$[H_\nabla + H_{ne} + aH_{ee}, S_{op}^2 \text{ or } S_{op,z}] = 0, \qquad (Eq.17)$$

what we have extended with the coupling strength parameter "a"; index "op" stands for "operator". Consequently, the exact eigenfunctions, not single determinant $\Psi_k$ of Eq.1 (a=1) or single determinant $Y_k$ of Eq.2 (a=0) are also eigenfunctions of the two spin operators [1], which in the case of a=0

$$S_{op}^2 Y_k = S(S+1)Y_k , \qquad (Eq.18)$$
$$S_{op,z} Y_k = M_S Y_k , \qquad (Eq.19)$$

where S and $M_S$ are the spin quantum numbers describing the total spin $S = \Sigma_{i=1...N} s_i$ and its z component of an N electron (TNRS, a=0) state $Y_k$. For example, let us suppose that one is interested in the singlet states of a molecule. In this case those $Y_k$ determinants must be eliminated from the determinant expansion of which $M_S \neq 0$, i.e., which are not singlets, - just like in the routine HF-SCF/basis/a=1 calculations, for obtaining the lowest lying triplet/singlet energy of an atom via manipulating the input multiplicity for $S_0$. The spin algebra for Eqs.14 and 16 is exactly the same as for basis set $\{S_k\}$, described in ref.[1], and not repeated here for the sake of brevity. Only the two simplest spin-adapted cases are mentioned when N=even in $Y_0$ obtained from HF-SCF/basis/a=0, the doubly excited singlet $Y_{p(\alpha)p(\beta)}^{r(\alpha)r(\beta)}$, wherein $(\alpha,\beta)$ electron pair from p orbital below LUMO are promoted to r orbital over HOMO with the same $(\alpha,\beta)$ spin configuration as indicated in brackets, as well as the singly excited singlet configuration:

$$2^{-1/2}(Y_{p(\alpha)}^{r(\alpha)} + Y_{p(\beta)}^{r(\beta)}) . \qquad (Eq.20)$$

Notice that in Eq.20 both terms alone are also diagonal elements in Eq.15, but not pure spin states.

## 6.b. A quick power series estimation for the ground state (k=0) of SE (Eq.1, a=1) from TNRS (Eq.2, a=0) starting from $E_{electr,0}$(TNRS) in Eq.15

The quick estimation to convert Eq.2 to Eq.1 commences from Eq.15, because it accounts for the big part of the large difference in energy value between Eqs.1 and 2, as Figure 2 demonstrates with the solid line. Importantly, it is the result of the simplest (1x1) CI matrix. One can use larger CI matrix, or alternatively a quick method, e.g. the "moment expansion" of the electron density (Appendix 2). Based on these, here we extend the form in Eq.15 as

$$E_{electr,0} \approx E_{electr,0}(\text{TNRS with L}^{th} \text{ order PS expansion}) \equiv e_{electr,0} + \Sigma_{j=1...L}(a_j t^j + b_j v_{ne}^j + c_j z^j) =$$
$$= E_{electr,0}(\text{TNRS}) + (a_1 t + b_1 v_{ne} + (c_1-1)z) + \Sigma_{j=2...L}(a_j t^j + b_j v_{ne}^j + c_j z^j) , \qquad (Eq.21)$$

where the pre-calculated $t \equiv <Y_0|H_\nabla|Y_0>$, $v_{ne} \equiv <Y_0|H_{ne}|Y_0>$ and $z \equiv a<Y_0|H_{ee}|Y_0>$ integrals are used, as well as $c_1 z = z + (c_1-1)z$ being used to show its more visible extension from Eq.15. PS stands for power series. In Eq.15 and Eq.21 the most important a=1 is involved. In Eq.21 if $a \neq 1$, then $enrg_{electr,0}(a)$ replaces $E_{electr,0}(a=1)$, and if a=0 then $a_j = b_j = c_j = 0$ for all j, as well as coupling strength parameter "a" is not to be confused with PS coefficients $a_j$. The MP perturbation uses the excited states $(Y_k)$ in the expansion ("vertical" algebraic way), while Eq.21 uses only $Y_0$ ("horizontal" algebraic way), notice

that the latter is practically instant in respect to computation, while the former can be time consuming.

We have obtained the coefficients in Eq.21 by least square fitting to 149 ground state G3 molecular energies (Figure 1) to minimize the average absolute deviation. For k=0, Eq.15 (with HF-SCF/STO-3G/a=0) is greatly improved by Eq.21 (with HF-SCF/STO-3G/a=0 and L= 2 and 3), most importantly the latter is better than the HF-SCF/STO-3G/a=1 regular calculation, so the correlation effect is somehow accounted for by Eq.21. We emphasize that in this work no more serious correlation calculation is considered than Eq.21, we just want to demonstrate the way to use Eq.2 for solving Eq.1. We have obtained the following values for PS coefficients in Eq.21 for the second order (L=2) case as

```
a₁= -0.761233,      b₁= -0.448435,      c₁= 0.430207
a₂=  2.270220E-004, b₂= -5.068453E-005, c₂= 1.678742E-004
```
, while the third order (L=3) PS coefficients are
```
a1= -0.853118,      b1= -0.519268,      c1=  0.289831
a2=  5.224182E-004, b2= -1.321651E-004, c2=  6.744563E-004
a3= -2.026111E-007, b3= -2.221198E-008, c3= -4.823247E-007.
```
The average absolute deviation in h and % from G3 data are:

```
HF-SCF/STO-3G/a=1: 3.497650 h or 1.88 %,
L=2 in Eq.21      : 1.615905 h or 1.02 %,
L=3 in Eq.21      : 1.563234 h or 1.06 %.
```

As can be seen, the regular HF-SCF/basis/a=1 calculation is improved, so correlation effects are accounted for by Eq.21, as well as the L=3 level not improving much over L=2 ("small power" property, see Appendix 2). Larger L can yield not-realistic values for coefficients, a known problem that can happen in least square fit with PS expansion. The latter means that, e.g. the L= 2 and 3 level coefficients are realistic in that they correct the different energies with main terms, which are the j=1 terms in Eq.21: Negative $a_1$ and $|a_1|<1$ necessary to subtract a part of kinetic energy away, so for $b_1$ to keep the virial theorem hold, as should $0<c_1<1$ be also.

## 7. Generalization of Brillouin's theorem in relation to coupling strength parameter

It is interesting to consider the Brillouin's theorem (1934) [1, 30] in this respect. To avoid becoming lost in the jungle of notations, we summarize, or list the notations from above as initial conditions for generalized Brillouin's theorem: $y_k(a)$ is the exact $k^{th}$ excited state solution of Eqs.1 and 2, extended with coupling strength parameter "a" knowing that $y_k(a=0)$ have a single determinant form while $y_k(a\neq0)$ have non-single determinant form, $Y_k \equiv y_k(a=0)$ can be obtained via the HF-SCF/basis/a=0 algorithm for any k; $\Psi_k \equiv y_k(a=1)$ is the physical wave function for ground (k=0) and excited (k>0) states of molecular systems (one of the ultimate goal in computation chemistry), as well as HF-SCF/basis/a=1 (the regular HF-SCF/basis) algorithm provides the famous approximation $S_0 \approx \Psi_0$, along some lowest lying excited states $S_k$, called N-electron Slater determinants. Now, we discuss the $s_0(a)$ single Slater determinant HF-SCF/basis/a calculation, of which $s_0(a) \approx y_0(a)$, and particularly, $y_0(a=0) = s_0(a=0) \equiv Y_0$ and $\Psi_0 \approx s_0(a=1) \equiv S_0$ holds. Starting from lowest lying state $Y_0$ or $S_0$, one can make singly excited Slater determinant [1] basis elements to describe $\Psi_k$ for lower k values, most importantly for k=0 by replacing a spin-orbital HOMO level or below (call it b) to a spin-orbital LUMO level or higher (call it r), denoted as $Y_{0,b}^r$ and $S_{0,b}^r$, respectively. (For example, in $Y_0 = |\alpha_1 f(\mathbf{r}_1), \beta_2 f(\mathbf{r}_2), \alpha_3 g(\mathbf{r}_3)>$, $Y_{k'} = |\alpha_1 f(\mathbf{r}_1), \alpha_2 g(\mathbf{r}_2), \beta_3 g(\mathbf{r}_3)>$, and $Y_{k''} = |\alpha_1 f(\mathbf{r}_1), \alpha_2 g(\mathbf{r}_2), \alpha_3 h(\mathbf{r}_3)>$ with $e_{electr, 0 \text{ or } k' \text{ or } k''} = 2\varepsilon_1 + \varepsilon_2$, $\varepsilon_1 + 2\varepsilon_2$, and $\varepsilon_1 + \varepsilon_2 + \varepsilon_3$ respectively, the three, f, g and h, MOs split into six spin-orbitals ($\alpha f, \beta f, \alpha g, \beta g, \alpha h, \beta h$) in counting columns in the basis set elements $Y_k$ and analogously $S_k$ – a known method.) Brillouin's theorem states that $<S_0|H|S_{0,b}^r> = 0$ as a consequence of the HF-SCF/basis/a=1 algorithm [1]. For this reason, extending $S_0$ with only singly excited determinants to improve for $\Psi_0$ or improve for $\Psi_0$ and estimate $\Psi_1$ is impossible, the doubly excited determinants $S_{0,bc}^{rs}$ are necessary and are the most important corrections to $\Psi_0$, more exactly the $\{S_0, \{S_{0,b}^r\}, \{S_{0,bc}^{rs}\}\}$ basis set. (Although these Brillouin matrix elements are zero, the singly excited $S_{0,b}^r$ do have an effect on $\Psi_0$ via Hamilton matrix elements as $<S_{0,b}^r|H|S_{0,bc}^{rs}>$.) With the language of linear algebra, the $s_0(a \neq 0)$ approximate solution in this integral product does what eigenfunctions can do typically: annihilates the operator in the core and the product has an exact eigenvalue, particularly zero.

A trivial extension of Brillouin's theorem for cases HF-SCF/basis/a (which approximates $y_0(a)$ by single determinant $s_0(a)$) is formally the same, that is

$$<s_0(a)|H_\nabla + H_{ne} + aH_{ee}|s_{0,b}^r(a)> = 0 , \qquad (Eq.22)$$

and the proof is the same. Eq.22 for a=0, i.e. for Eq.2 and its generated $\{Y_k\}$ eigenfunction set (as a newly introduced candidate basis set for CI treatment for Eq.1) tells us only that the triviality such as $<Y_0|H_\nabla + H_{ne}|Y_{0,b}^r> = 0$, although the more general $<Y_{k'}|H_\nabla + H_{ne}|Y_k> = 0$ is also true of Eq.16 with a=0 for k'≠k, where indices k and k' count the ground ($Y_0$), singly ($Y_{0,b}^r$), doubly ($Y_{0,bc}^{rs}$), … n-touply excited Slater determinants as well, because $Y_k$'s are eigenfunctions. Like Hund's rule annihilates at a=0 (not detailed here), Brillouin's theorem becomes a triviality, because $s_0(a=0)$ becomes equal to $Y_0$, that is, an approximate form becomes an exact form. Eq.22 for eigenvalues trivially yields $<y_{k'}(a)|H_\nabla + H_{ne} + aH_{ee}|y_k(a)> = 0$ also for the wider range k'≠k, because $y_k(a)$'s are eigenfunctions. The Brillouin theorem (the original, wherein a=1 in Eq.22) and its extension (here, with a≠1 in Eq.22), wherein "a" can be any in Eq.22 tells us more, because $s_0(a)$ and $s_{0,b}^r(a)$ are not eigenfuntions of $H_\nabla + H_{ne} + aH_{ee}$, yet (and this is the point in Brillouin theorem) these matrix elements are still zero, a characteristic property from the HF-SCF/basis/a algorithm. As in the discussion on Eq.20 above, the right hand side of Eq.16 is zero if a=0, or zero if $Y_{k'}$ or $Y_k$ differ in three or more spin-orbitals. For example, with a≠0 in Eq.16, the $Y_{0,bcde}^{rspq}$ and $Y_{0,bcde}^{rsvw}$ differ in only two spin-orbitals, and do not yield zero for the right hand side of Eq.16 or the left hand side of Eq.20. In this way, Eq.22 reduces to sub-cases of Eq.16 if a=0, but Eq.16 with a≠0 tells us even more than Eq.22, the reason being that the operator in Eq.22 and in wave functions have the same "a" values, while in Eq.16 the operator contains a value of "a", but the wave function is $Y_k \equiv y_k(a=0)$ for k and k' i.e., two different "a" values are involved. An important consequence of this is that, for Eq.1, the $\{Y_0, \{Y_{0,b}^r\}\}$ truncated basis set generated by Eq.2 (using the minimal, singly-excited ones) can already be used as a basis to estimate $\Psi_0$ better than e.g., Eq.8, even to estimate $\Psi_1$ also by the eigenvectors of the Hamiltonian matrix. This means that, it can provide the large part of correlation energy, and the doubly excited determinants do not have to be calculated to save computer time and disc space unless one needs more accurate results or higher excited states. Again, in the literature the CI calculation is based on HF-SCF/basis/a=1 generated $\{S_0, \{S_{0,b}^r\}, \{S_{0,bc}^{rs}\}\}$ or a higher basis set to solve Eq.1, while here, we are talking about the HF-SCF/basis/a=0 generated $\{Y_0, \{Y_{0,b}^r\}\}$ or higher basis set to solve Eq.1.

It is important to mention that, to stop at single excited determinants in set $\{Y_k\}$ can be restricted by e.g., symmetry reason. For example, even the simple $H_2$ molecule (p.63 of ref.[1]) owns the property that (gerade HOMO and ungerade LUMO) restricts the wave function approximation by excluding single excited determinants (that is, producing zero CI Hamiltonian matrix elements). It yields the $\Psi_0 \approx c_0 Y_0 + c_{12}^{34} Y_{12}^{34}$ simplest improvement over $c_0 Y_0$ e.g. in basis $\{Y_k\}$ as is well known for $S_0$ and $S_{12}^{34}$, i.e., the simplest CI approximation needs double excited determinant.

**Conclusions**

The coupling strength parameter extended Hamiltonian $H(a) = H_\nabla + H_{ne} + a H_{ee}$ has been analyzed in relation to the mathematical TNRS (a=0) and the physical (a=1) cases, as well as algorithms have been outlined on how to convert the electronic energy from case a=0 to a=1 in computation. The Brillouin theorem has been extended from the original a=1 to general a≠1 cases, along with discussing its consequences on determinant basis sets in CI calculations. Important also is the extension of 1st Hohenberg-Kohn theorem as $Y_0(a=0) \Leftrightarrow H_{ne} \Leftrightarrow \Psi_0(a=1)$ manifesting as $E_{electr,0} = e_{electr,0} + \langle\Psi_0|H_{ee}|Y_0\rangle/\langle\Psi_0|Y_0\rangle$. The correct single determinant form solutions $\{Y_k(a=0)\}$ suffering from basis set error only, can always be obtained in one step by HF-SCF/basis/a=0 in contrast to the not correct single determinant form $S_0(a=1)$ from HF-SCF/basis/a=1 in relation to the non-single determinant $\Psi_0$, which always needs more steps to converge. (The benefit of the latter is well known, the former is introduced as an alternative.) The $\{Y_k(a=0)\}$ provides us a mathematically correct, ortho-normal, well-behaving basis set for CI calculations, and the single excited determinants from $Y_0$ have more freedom than that which is restricted by Brillouin theorem wherein $S_0$ is used to generate basis set for CI. The first approximation $E_{electr,k} \approx E_{electr,k}(TNRS) \equiv e_{electr,k} + (N(N-1)/2)\langle Y_k|r_{12}^{-1}|Y_k\rangle$ (coming from trial function via the variation principle for k=0) for ground (k=0) and excited (k>0) states has been well founded via (extending the MP perturbation theory also, and) the linear algebra of the CI theory, but of course finer correlation calculation (DFT, MP, etc.) or eigensolving certain levels of the CI matrix is necessary for chemical accuracy. This expression constitutes the diagonal elements of the TNRS-CI matrix, while the off-diagonal elements are simply the $\langle Y_{k'}|H_{ee}|Y_k\rangle$, (simpler in contrast to off-diagonal elements generated by basis set based on $S_0$).

## Appendix 1

Further relation to Eqs.6-9 is the obvious $E_{electr,k} > e_{electr,k}$, because $1/r_{ij} \geq 0$ always, but for the sake of chemical accuracy (1 kcal/mol), the $E_{electr,0} \gg e_{electr,0}$ is more plausible algebraically. For even further relations, one can start from the variation principle: Let the normalized solution of Eq.2, the $Y_0$, be a trial for Eq.1, and one gets $E_{electr,0} \leq \langle Y_0|H|Y_0\rangle = \langle Y_0|H_{ee}|Y_0\rangle + \langle Y_0|H_\nabla+H_{ne}|Y_0\rangle = \langle Y_0|H_{ee}|Y_0\rangle + e_{electr,0}$, that is

$$E_{electr,0} \leq e_{electr,0} + \langle Y_0|H_{ee}|Y_0\rangle . \qquad (Eq.A1)$$

The reverse situation, when $\Psi_0$, the solution of Eq.1 is a trial function for Eq.2, one gets the simpler

$$e_{electr,0} \leq \langle \Psi_0|H_\nabla+H_{ne}|\Psi_0\rangle . \qquad (Eq.A2)$$

Equality holds for both in Eqs.A1-A2 in the trivial case N=1, because there $\Psi_0(N=1) = Y_0(N=1)$. From Eq.1, it separates as $\langle \Psi_0|H|\Psi_0\rangle = \langle \Psi_0|H_\nabla+H_{ne}|\Psi_0\rangle + \langle \Psi_0|H_{ee}|\Psi_0\rangle = E_{electr,0}$, and the right hand side is majored by Eq.A1 as $\langle \Psi_0|H_\nabla+H_{ne}|\Psi_0\rangle + \langle \Psi_0|H_{ee}|\Psi_0\rangle \leq e_{electr,0} + \langle Y_0|H_{ee}|Y_0\rangle$, and with Eq.A2 one obtains

$$\langle \Psi_0|H_{ee}|\Psi_0\rangle \leq \langle Y_0|H_{ee}|Y_0\rangle . \qquad (Eq.A3)$$

The counterpart of Eq.A1 comes from Eq.A2 with an extension as $e_{electr,0} + \langle \Psi_0|H_{ee}|\Psi_0\rangle \leq \langle \Psi_0|H_\nabla+H_{ne}|\Psi_0\rangle + \langle \Psi_0|H_{ee}|\Psi_0\rangle = \langle \Psi_0|H|\Psi_0\rangle = E_{electr,0}$ which is

$$E_{electr,0} \geq e_{electr,0} + \langle \Psi_0|H_{ee}|\Psi_0\rangle . \qquad (Eq.A4)$$

In summary the full relation is

$$e_{electr,0} \ll (e_{electr,0}+\langle \Psi_0|H_{ee}|\Psi_0\rangle) \leq$$
$$E_{electr,0} = (e_{electr,0}+\langle \Psi_0|H_{ee}|Y_0\rangle/\langle \Psi_0|Y_0\rangle) \leq (e_{electr,0}+\langle Y_0|H_{ee}|Y_0\rangle) \quad (Eq.A5)$$

which extends Eq.A3 as

$$\langle \Psi_0|H_{ee}|\Psi_0\rangle) \leq \langle \Psi_0|H_{ee}|Y_0\rangle/\langle \Psi_0|Y_0\rangle \leq \langle Y_0|H_{ee}|Y_0\rangle . \qquad (Eq.A6)$$

One other expression stemming from the variation principle has to be emphasized: the $S_0$ from HF-SCF/basis/a=1 for Eq.1 is energetically better than $Y_0$ from HF-SCF/basis/a=0 for Eq.2, when one uses this $Y_0$ for Eq.1, that is,

$$E_{electr,0} < \langle S_0|H|S_0\rangle \leq e_{electr,0} + (N(N-1)/2)\langle Y_0|r_{12}^{-1}|Y_0\rangle , \qquad (Eq.A7)$$

where the equality may come up when small e.g., STO-3G basis set is used. Eq.A7 is an extension of Eq.A1. The error (called correlation energy) of the middle part with $S_0$ in Eq.A7 stems from the fact that $\Psi_0$ is approximated with incorrect wave function form, namely with $S_0$. The left part is a classically known relation (variation principle) in Eq.A7, while the expression on the right hand side for $Y_0$ comes from first perturbation and not from energy minimization, so the right side relationship between expressions containing $S_0$ vs. $Y_0$ comes from a variation principle, but at least the LCAO coefficients vary slowly between $Y_0(a=0)$ and $S_0(a=1)$. (Again, LCAO parameters in correct functional form $Y_0$ come from solving Eq.2 numerically, while in the incorrect functional form $S_0$ the LCAO parameters come from the energy minimization of $\langle S_0|H|S_0\rangle$ for Eq.1, restricted by the known ortho-normalization for MOs.)

## Appendix 2

Among moment expansions the most famous are the Thomas-Fermi ($T \approx c_F \int \rho_0^{5/3} d\mathbf{r}_1$) or Weizsacker ($T \approx (1/8)\int|\nabla_1\rho_0|^2/\rho_0 d\mathbf{r}_1$) approximations, Dirac formula ($V_{ee} \approx B_D \int \rho_0^{4/3} d\mathbf{r}_1$), as well as Parr terms $c_{AB}(\int\rho_0^A d\mathbf{r}_1)^B$ in the power series wherein easy formulas for A and B keeps it scaling correct up to infinity, separately for T and $V_{ee}$. Common in these formulas: 1, the first terms, come from plausible assumptions and derivations, but secondary and higher terms definitely necessary for chemical accuracy, (a manifest example is that only the Thomas-Fermi approximation for T fails to describe chemical bonds), 2, first terms with proper (but generally empirical) parameters can be used to account for the entire term T or $V_{ee}$, or just for their correction, depending (quite surprisingly) on how the user wants to define them, 3, a plausible series in principle converges to the accurate energy value, but as a drawback, probably coming from the imperfect parametrization, only small power terms (2 to 4) can account for a large pool of molecular systems, which, while increasing the power decrease the range of molecular systems in terms of accuracy.

**Main notations and definitions**

argument: If no argument, we mean the exact accurate value, e.g. $E_{electr,0}$ or $\rho_0$,
  if argument written, it may refer the method used, e.g. $E_{electr,0}$(CI or HF-SCF),
  or it may refer its important dependence in question, e.g. $\rho_0 = \rho_0(\mathbf{r}_1) = \rho_0(\mathbf{r}_1,a)$,
  but not necessarily all variables or parametric variables are written out in it.
AC= adiabatic connection
AO= atomic orbital
CI= configuration interactions
$\delta_{ij}$ = 1 if i=j and 0 if i≠j, the Kronecker delta
DFT= density functional theory
GTO= Gaussian type atomic orbital
$H_\nabla = -(1/2)\Sigma_{i=1,\ldots,N}\nabla_i^2$, the kinetic energy operator in a.u.
$H_{ne} = -\Sigma_{i=1,\ldots,N}\Sigma_{A=1,\ldots,M}Z_A R_{Ai}^{-1}$, the nuclear – electron attraction operator in a.u.
$H_{ee} = \Sigma_{i=1,\ldots,N}\Sigma_{j=i+1,\ldots,N}r_{ij}^{-1}$, the electron – electron repulsion operator in a.u.
$H(a) = H_\nabla + H_{ne} + aH_{ee}$ Hamiltonian for $H(a)y_k(a) = \text{energy}_{electr,k}(a)y_k(a)$, k=0,1,2,…,
  a= coupling strength parameter, a=1 makes non-relativistic physical sense only,
    the most important interval is [0,1], the HF-SCF/basis/a routine yields the
    energy optimized single determinant $s_0(a)$ approximation for $y_0(a)$,
  if a=0, then the exact form of $y_k$ is single determinant, otherwise not,
  a= 1 case: $(H_\nabla + H_{ne} + H_{ee})\Psi_k = E_{electr,k}\Psi_k$, $\rho_k(\mathbf{r}_1,a=1)=\int\Psi_k^*\Psi_k ds_1 d\mathbf{x}_2\ldots d\mathbf{x}_N$, see Eq.1,
    $S_0$= normalized ($<S_0|S_0>$=1) single determinant via HF-SCF/basis/a=1 routine
        approximating $\Psi_0$ via energy minimizing $E_{electr,0}$(HF-SCF)$\equiv <S_0|H(a=1)|S_0>$,
        suffering from basis set and correlation error: $E_{electr,0}$(HF-SCF)$< E_{electr,0}$,
    $y_0(a=1) = \Psi_0$, $\rho(\mathbf{r}_1,\text{HF-SCF/basis/a}=1)=\int S_0^*S_0 ds_1 d\mathbf{x}_2\ldots d\mathbf{x}_N$,
    $\{S_k\}$= single, double, etc. excited Slater determinant from $S_0$ with using
        LUMO+1,+2, etc., it is ortho-normalized basis set.
  a= 0 case: $(H_\nabla + H_{ne})Y_k = e_{electr,k}Y_k$, $\rho_k(\mathbf{r}_1,a=0)=\int Y_k^*Y_k ds_1 d\mathbf{x}_2\ldots d\mathbf{x}_N$, see Eq.2,
    $Y_k$= is a normalized ($<Y_k|Y_k>$=1) and single determinant form exact
        solution of Eqs.2 or 4, suffering from basis set error via HF-SCF/basis/a=0,
    $y_0(a=0) = Y_0$. $E_{electr,k}$(TNRS)$\equiv e_{electr,k} + <Y_k|H_{ee}|Y_k>$,
    $\{Y_k\}$= ortho-normalized complete basis set.
HF-SCF= Hartree- Fock self consistent field method

HF-SCF/basis/a= conventional HF-SCF/basis routine for estimating ground state
$H(a)y_0(a)=$ energy$_{electr,0}(a)$ with additional programming input for the value of "a"
HK= Hohengerg – Kohn (theorem, 1$^{st}$ and 2$^{nd}$)
KS= Kohn – Sham (algorithm)
LCAO= linear combination of atomic orbitals
M= number of atoms in the molecular system
MO= molecular orbital
MP= Møller–Plesset perturbation theory
N= number of electrons in the molecular system
PES= potential energy surface
RPA = random phase approximation
S= $\Sigma_{i=1…N}s_i$= total spin as the sum of individual electron spins, multiplicity= 2S+1, not to be confused with the S$_0$ single determinant (heavily used here) or the term symbols (not used here)
SE= non-relativistic electronic Schrödinger equation
STO= Slater type atomic orbital
t or v$_{ne}$≡ <Y$_0$|H$_\nabla$ or H$_{ne}$|Y$_0$>= ground state kinetic or nuclear-electron attraction energy of TNRS (a=0), respectively
T or V$_{ne}$ or V$_{ee}$≡ <Ψ$_0$|H$_\nabla$ or H$_{ne}$ or H$_{ee}$|Ψ$_0$>= ground state kinetic or nuclear-electron attraction or electron-electron repulsion energy of real (a=1) system
TNRS= totally non-interacting reference system, see Eq.1, a=0 annihilates H$_{ee}$ in H
V$_{nn}$= nuclear-nuclear repulsion energy
well behaving: Y$_k$ is well behaving via Eq.2 for any molecular frame H$_{ne}$, only the zero trivial solution has not-finite integral. Well behaving property is important to ask for S$_0$, which can never approximate Ψ$_0$ correctly with single-determinant.

For easier reading, most importantly: The (y$_k$(a),enrg$_{electr,k}$(a)) is the k-th eigenvalue pair of electronic Hamiltonian H(a), most importantly, we use distinguishing notations for a=0 (Eq.2) as (Y$_k$,e$_{electr,k}$) and for a=1 (Eq.1) as (Ψ$_k$,E$_{electr,k}$), as well as S$_0$ (generally s$_0$(a)) is a single determinant approximation for Ψ$_0$ (generally for y$_0$(a)) via HF-SCF/basis/a=1 (generally with a) energy minimizing algorithm, (k=0,1,2,…, enrg$_{electr,k}$(a)≤ enrg$_{electr,k+1}$(a)). As analyzed, y$_k$(a=0)= Y$_k$ has a single determinant form solution, while y$_k$(a≠0) do not, and E$_{electr,0}$(method) approximates E$_{electr,0}$ by a certain method (HF-SCF, KS, G3, CI, etc.).

**Acknowledgments**

Thanks to Prof. Istvan Mayer at CRC-HAS for his short but very essential guiding where the Monstergauss 1981 had to be modified for the purpose of this work. Financial support for this research from OTKA-K 2015-115733 and 2016-119358 is kindly acknowledged.

**Figures:**

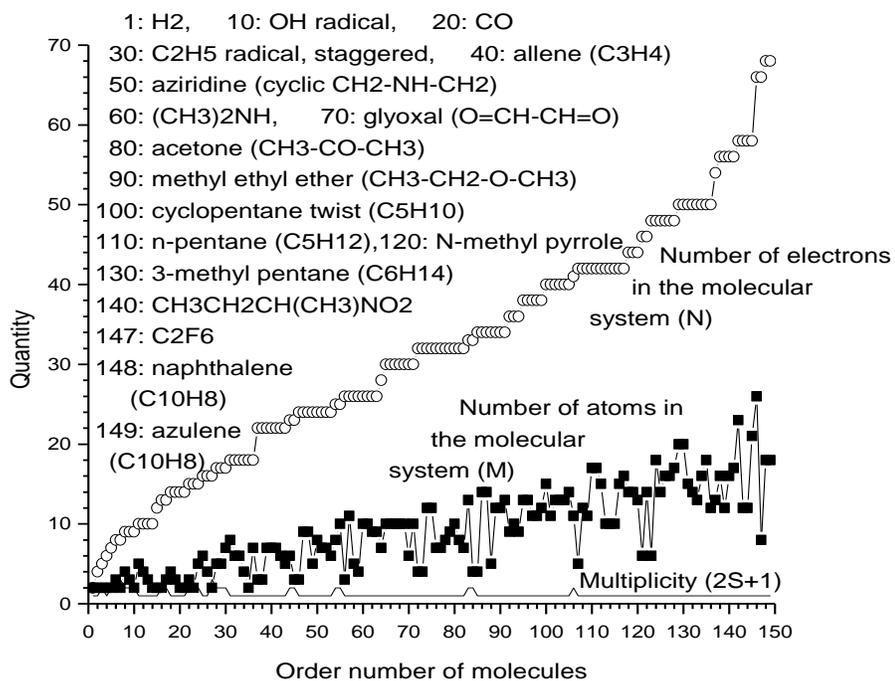

Figure 1.: Some data of the nuclear frame of 149 neutral molecules from the G3 set for Figure 2 to identify them in our analysis.

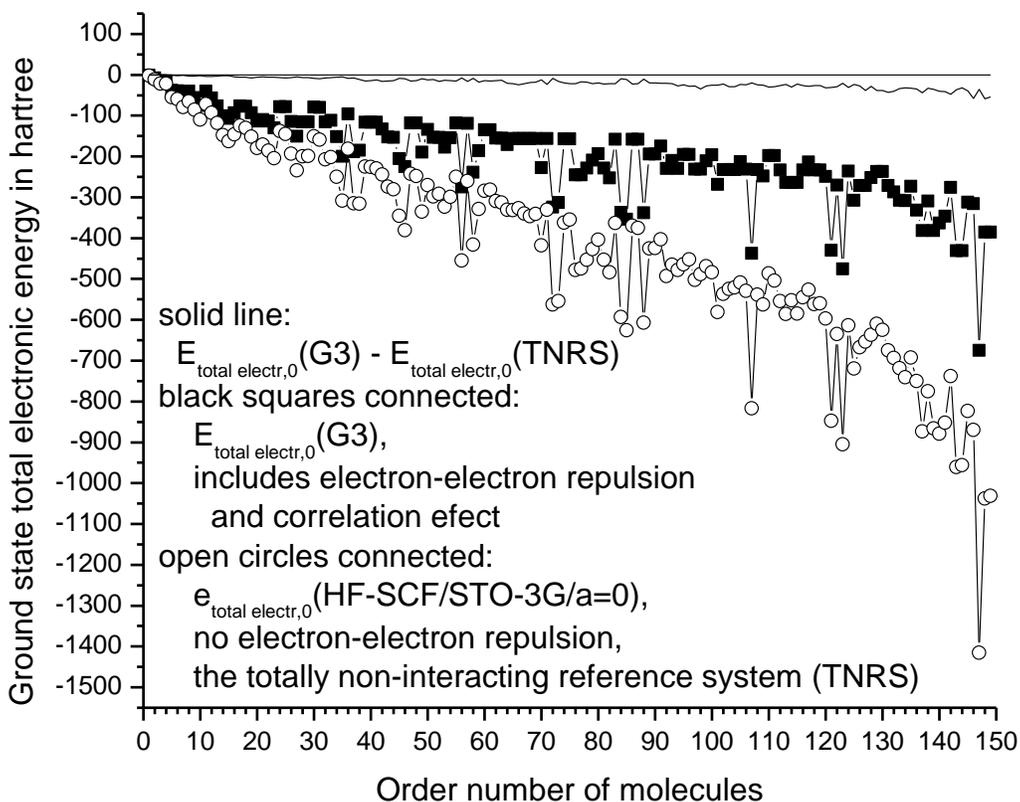

Figure 2.: Ground state total electronic energy of molecules is plotted as function of the order number of molecules shown on Figure 1. The order number is chosen for number of electrons (N) to be monotonic, so local peaks come from very different molecules having same (or close) N values but different ground state total electronic energies, i.e. the shape of the curve itself has no particular meaning, the important message is that the two curves (black squares and open circles) run together like the same fingerprint. The related curve for open circles with larger 6-31G** basis set would yield lower energy values by about 2% (basis set error improvement), and would be almost at the same position for eyes, not plotted. Solid line is the deviation $E_{total\ electr,0}$(G3) - $E_{total\ electr,0}$(TNRS) via first approximation in Eq.8 which brings the open circle values (Eq.2 with small basis set error) remarkably back to black square ones (Eq.1 with G3 estimation), it is also the approximate error of the (1,1) element of the CI matrix in Eq.14.